\documentclass[superscriptaddress,showkeys,prl,aps,twocolumn]{revtex4-1}
\usepackage{dcolumn}
\usepackage{bm}

\usepackage[colorlinks,hyperindex, urlcolor=blue, linkcolor=blue,citecolor=black, linkbordercolor={.7 .8 .8}]{hyperref}
\usepackage{graphicx}
\usepackage{float}
\usepackage{amsfonts}
\usepackage{amsmath}
\usepackage{amssymb}
\usepackage{amsbsy}
\usepackage{nicefrac}

\begin{document}

\title {Signatures of phonon and defect-assisted tunneling in planar metal-hexagonal boron nitride-graphene junctions}

\author{U. Chandni}
\email{chandniu@gmail.com}
\address{Institute for Quantum Information and Matter, Department of Physics, California Institute of Technology, 1200 E. California Blvd., Pasadena, California 91125, USA}
\author{K. Watanabe}
\address{National Institute for Materials Science, 1-1 Namiki Tsukuba, Ibaraki 305-0044, Japan}
 \author{T. Taniguchi}
\address{National Institute for Materials Science, 1-1 Namiki Tsukuba, Ibaraki 305-0044, Japan}
\author{J. P. Eisenstein}
\address{Institute for Quantum Information and Matter, Department of Physics, California Institute of Technology, 1200 E. California Blvd., Pasadena, California 91125, USA}

\begin{abstract}
Electron tunneling spectroscopy measurements on van der Waals heterostructures consisting of metal and graphene (or graphite) electrodes separated by atomically thin hexagonal boron nitride tunnel barriers are reported.  The tunneling conductance $dI/dV$ at low voltages is relatively weak, with a strong enhancement  reproducibly observed to occur at around $|V|\sim$ 50 mV.  While the weak tunneling at low energies is attributed to the absence of substantial overlap, in momentum space, of the metal and graphene Fermi surfaces, the enhancement at higher energies signals the onset of inelastic processes in which phonons in the heterostructure provide the momentum necessary to link the Fermi surfaces.  Pronounced peaks in the second derivative of the tunnel current, $d^2I/dV^2$, are observed at voltages where known phonon modes in the tunnel junction have a high density of states.  In addition, features in the tunneling conductance attributed to single electron charging of nanometer-scale defects in the boron nitride are also observed in these devices. The small electronic density of states of graphene allows the charging spectra of these defect states to be electrostatically tuned, leading to ‘Coulomb diamonds’ in the tunneling conductance.

\end{abstract}

\date{\today}

\maketitle
\section{Introduction}
Tunneling spectroscopy has proved to be an excellent tool for studying low-dimensional electronic systems \cite{Smoliner}. In addition to revealing features pertaining to the density of states (DOS) of the junction electrodes~\cite{Reed}, tunneling can probe the geometry of the electrode Fermi surfaces~\cite{JPE_PRB91,Gennser_PRL}, electron-electron interactions both within and between the electrodes~\cite{JPE_PRL92}, phonon energy scales~\cite{Chynoweth}, defects within the tunnel barrier~\cite{Chandni_NL}, etc.  While scanning tunneling microscopy (STM) is extensively used as a sensitive tool to detect the local DOS~\cite{Chen_STM}, macroscopic planar tunnel junctions provide averaged properties over larger length scales, often demanding reliable and reproducible device fabrication methods.  In the former, tunneling occurs at the atomic tip of the STM probe, thus involving electrons with a wide range of transverse momenta. In contrast, smooth planar tunnel junctions offer the possibility of momentum-resolved tunneling spectroscopy~\cite{Reed,JPE_PRB91,Gennser_PRL,Smoliner_PRL}.
\begin{figure*}[t]
\centering
\includegraphics[scale=0.27]{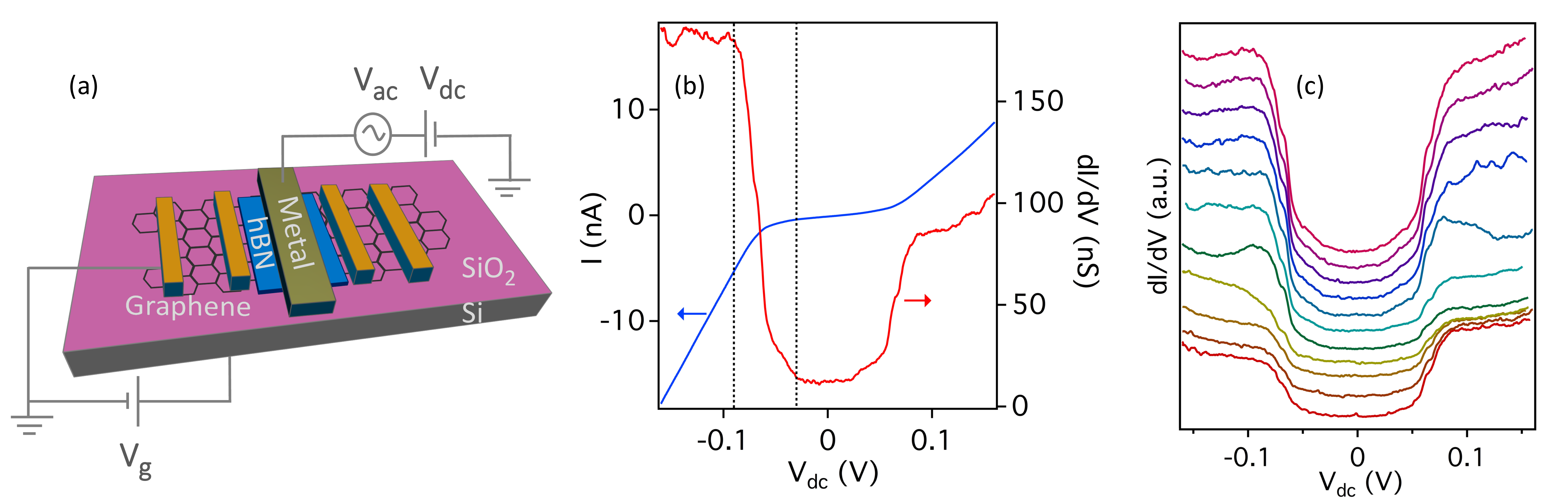}
\caption{(a) A general schematic of our device structure with SiO$_2$/Si acting as the substrate and a thin hBN as the tunnel barrier (not to scale). Gate voltage is applied to the SiO$_2 (\sim$ 285  nm) dielectric layer. An $ac+dc$ bias is applied to the top metal tunnel electrode (Ag or Cr/Au) and the tunnel current is measured using one of the Ohmic contacts (Cr/Au) to the bottom graphene layer. (b) Tunnel current ($I$) and differential conductance $dI/dV$  as a function of source-drain bias ($V_{dc}$) for a monolayer graphene device at zero gate voltage. The dotted vertical lines are at $V_{dc} = -30$ mV and $-90$ mV (c) $dI/dV$ at every 10 volts from $V_g$ = +50 (bottom) to -50V (top). The curves are offset for clarity.} 
\end{figure*}

Recent advances in creating van der Waals heterostructures based on graphene and hexagonal boron nitride (hBN) have opened new avenues to create atomically smooth tunnel junctions~\cite{Britnell_Science}. hBN is an isomorph of graphene with $\approx 2\%$ lattice mismatch and has been the preferred substrate material for high mobility graphene devices~\cite{Wang_Science}. Its large insulating band gap of $\sim 5.9$ eV and atomically smooth layers make hBN a good tunnel barrier candidate, which can potentially aid in-plane momentum conservation. Previous studies on simple metal-insulator-metal junctions with hBN as the insulator showed a linear current-voltage ($IV$) characteristic around zero-bias~\cite{Chandni_NL,Lee_APL}. The observed exponential dependence of the conductance on the hBN thickness established hBN as a good crystalline tunnel barrier. However, it was also observed that intrinsic nanometer-scale defect states present in the hBN atomic layers act like quantum dots and contribute to the tunneling $IV$ characteristics via single-electron charging events~\cite{Chandni_NL}.  In related works on graphene-hBN-graphene junctions, features seen in the $IV$ characteristic were attributed to resonant processes conserving the in-plane momentum and chirality of the tunneling electrons~\cite{Mishchenko_nnano}.

In this work we investigate tunnel junctions in which 2-5 atomic layers of hBN  (thickness $\lesssim$ 2 nm) are sandwiched between a graphene (or graphite) electrode and a conventional metal counter-electrode. Several distinct features of our results stand out.  Most dramatically, in all devices studied the tunneling conductance $dI/dV$ exhibits an abrupt and pronounced enhancement at bias voltages in excess of roughly $\pm$50 mV.  Our measurements reveal this enhancement to arise from a series of sharp peaks in the second derivative of the tunnel current, $d^2I/dV^2$.  The voltage locations of these peaks coincide with specific phonon modes in the heterostructure and are insensitive to a voltage applied to a back gate underlying the tunnel junction.   At lower bias, below the onset of phonon-assisted tunneling, we find that the tunneling conductance does respond to a back gate voltage, with a minimum in the conductance observed when the graphene is close to the Dirac point.  Finally, sharp features arising from Coulomb charging of intrinsic defect states in the hBN barrier layer are also sometimes observed.  Due to incomplete screening by the graphene electrode, these defect signatures also respond to a back gate voltage, and the classic ``diamond pattern" associated with Coulomb blockade is seen.

\section{Experimental Details}
The present tunnel junctions consist of a mono- or bilayer graphene (MLG or BLG) or graphite lower electrode, a thin ($\leq5 $ atomic layers) hBN tunnel barrier, and a metallic (either Cr/Au or Ag) top electrode.  These junctions are assembled on a SiO$_2$ substrate or on a thick hBN layer mounted on SiO$_2$.  In all cases the 285 nm thick SiO$_2$ layer itself lies atop a p-doped Si wafer which serves as a back gate and provides mechanical support.

To create these van der Waals heterostructures graphene (or graphite) and hBN flakes are first mechanically exfoliated onto separate Si wafers (also overlaid with SiO$_2$).  For tunnel devices with SiO$_2$ as the substrate material, the mechanically exfoliated graphene/graphite layer acts as the bottom electrode.  For tunnel devices with hBN as the substrate material, the graphene/graphite layer is first transferred onto a thick hBN layer atop a SiO$_2$ substrate using the polymer stamp dry transfer technique described in Ref.11.   Next, the thin hBN tunnel barrier is similarly transferred on to the graphene/graphite.  Finally, the metallic top tunnel electrode is lithographically fabricated by depositing Cr/Au (5/120 nm) or Ag (120 nm). Lateral dimensions of the junctions vary from $\sim 2\times2$ to $6\times6$   
$\mu$m$^2$.

The electrical transport properties of the graphene layer are measured using additional Cr/Au electrodes. A schematic of a typical device is shown in Fig.1(a). Tunneling current-voltage ($IV$) and differential conductance $dI/dV$ characterictics are measured simultaneously using standard lock-in techniques ($V=V_{dc}+V_{ac}$, with $V_{ac}=0.5$ mV at 13 Hz). The $d^2I/dV^2$ measurements were also performed simultaneously using a second lock-in amplifier synchronized at twice the excitation frequency. All the measurements were done at $T=4.2$ K.

\section{Results} 
Figure 1(b) displays $IV$ and $dI/dV$ characteristics from a tunnel junction consisting of a monolayer graphene (MLG) lower electrode, a hBN tunnel barrier, and a Cr/Au top electrode, all assembled on SiO$_2$/Si.  As the figure makes clear, a small but finite tunneling conductance is observed at $V_{dc}=0$.  Around $|V_{dc}| \sim 30$ mV $dI/dV$ begins to rise rapidly before leveling off beyond about $|V_{dc}| \sim 90$ mV.  

Applying a back gate voltage $V_g$ to the doped Si substrate tunes the free carrier density and, given the Dirac-like spectrum of MLG, the electronic density of states in the MLG electrode.  Figure 1(c) shows that the strong finite voltage enhancement of $dI/dV$ is unaffected by the back gate voltage, at least for $-50 \le V_g \le +50$ V. In contrast, the magnitude of the tunneling conductance around $V_{dc}=0$ does depend on the back gate voltage.  This is illustrated in Fig. 2(a) where the blue dots give the average tunneling conductance $\langle dI/dV \rangle$ over the range $|V_{dc}| \le 20$ mV while the red trace shows the measured resistivity $\rho_{MLG}$ of the graphene sheet.  Although not precisely coincident, the minimum in $\langle dI/dV \rangle$ occurs at nearly the same gate voltage as the maximum in the graphene resistivity.  Hence, consistent with expectations, the tunneling conductance is minimized at the Dirac point where the graphene density of states is lowest and its resistivity highest.  Figure 2(b) shows the tunneling conductance $dI/dV$ as a function of both $V_g$ and $V_{dc}$ on a logarithmic color scale. The finite voltage enhancement of $dI/dV$ appears as a vertical blue-colored band of uniform width, while the dark blue spot near $V_g=20$ V locates the graphene Dirac point.

\begin{figure}[t]
\centering
\includegraphics[scale=0.45]{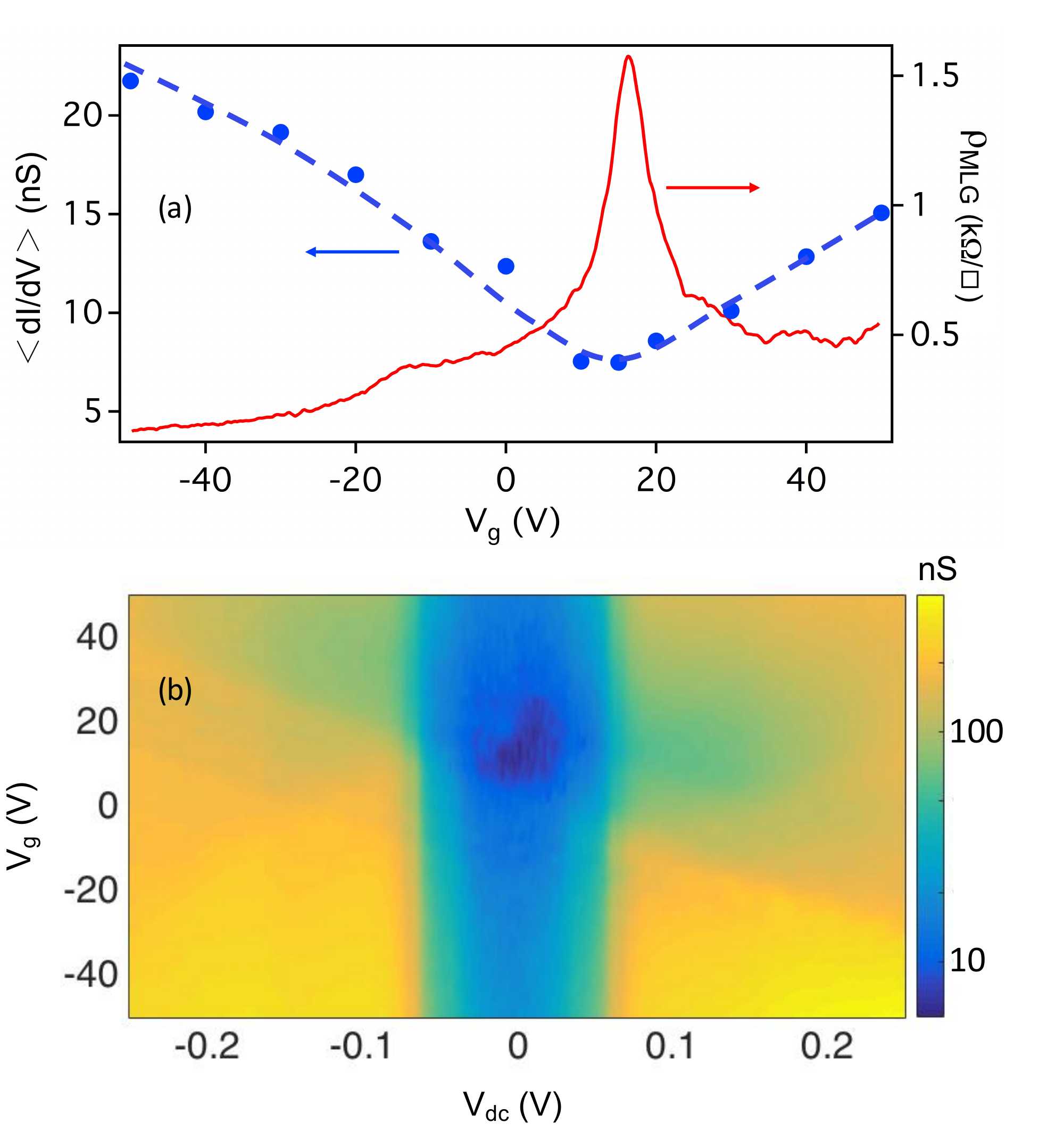}
\caption{(a) Average conductance around zero-bias  $ \langle dI/dV \rangle$ (blue) as a function of back-gate voltage ($V_g$) for the MLG device in Fig.1(b)-(c). The dashed line is a guide to the eye. The corresponding graphene resistivity ($\rho_{MLG}$) vs back-gate voltage is plotted in red. (b) A colormap of the differential conductance $dI/dV$ plotted in a log scale, as a function of the back gate voltage and source-drain bias.} 
\end{figure}

In order to elucidate the origin of these features in the tunneling characteristics, additional devices were fabricated, with variations in the tunnel electrodes and the underlying substrate material. Figure 3(a)-(c) show tunnel characteristics for three such devices consisting of (a) a MLG-hBN-Cr/Au tunnel junction, assembled on hBN substrate (b) BLG-hBN-Ag tunnel junction, assembled on SiO$_2$/Si substrate and (c) graphite-hBN-Cr/Au tunnel junction assembled on SiO$_2$/Si substrate respectively. As is evident, the abrupt strong enhancement of the tunneling conductance at finite voltage is observed in all the devices measured, independent of the substrate, the type of graphene/graphite electrode, and the metal forming the top tunnel electrode. Similar signatures were observed in 14 tunnel junctions from 8 devices. 

\begin{figure}
\centering 
\includegraphics[scale=0.4]{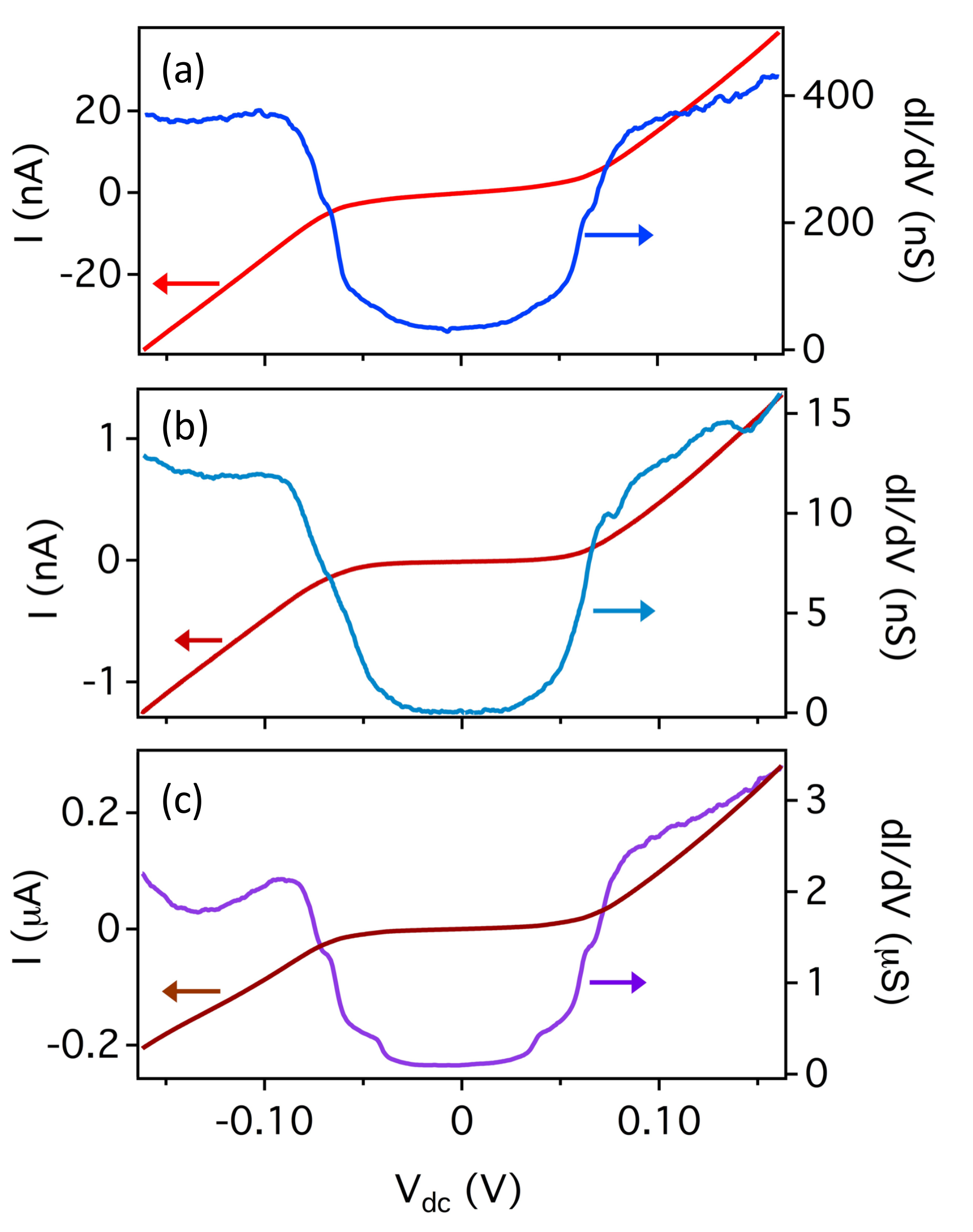}
\caption{Tunnel current ($I$) and differential conductance ($dI/dV$)  as a function of source-drain bias ($V_{dc}$) for (a) a MLG-hBN-Cr/Au tunnel junction on hBN substrate (b) BLG-hBN-Ag tunnel junction on SiO$_2$/Si substrate and (c) graphite-hBN-Cr/Au on SiO$_2$/Si substrate.}
\end{figure}

As the data in Fig. 3 suggests, the rapid enhancement of the tunnel conductance at finite bias occurs in a series of steps, better resolved in some junctions than in others.  To examine these features more closely, the second derivative of the tunnel current, $d^2I/dV^2$, was recorded.  Figure 4 shows typical results, with Fig. 4(a) showing the $d^2I/dV^2$ spectrum from a graphite-hBN-Cr/Au tunnel junction and Fig. 4(b) the spectrum from a MLG-hBN-Cr/Au device.  As is apparent, the spectra are virtually identical and show no gate voltage dependence (Fig.4(c)).  Four prominent peaks, labeled A, B, C, and D in the figure, are seen in each polarity.  Averaged over six devices, the voltage location of these peaks are $|V_A| = 36 \pm 3$ mV, $|V_B| = 61 \pm 2$ mV, $|V_C| = 74 \pm 2$ mV, and $|V_D| = 166 \pm 8$ mV.  As discussed below, we associate these peaks with phonon-assisted inelastic tunneling processes.

\begin{figure}[t]
\centering
\includegraphics[scale=0.45]{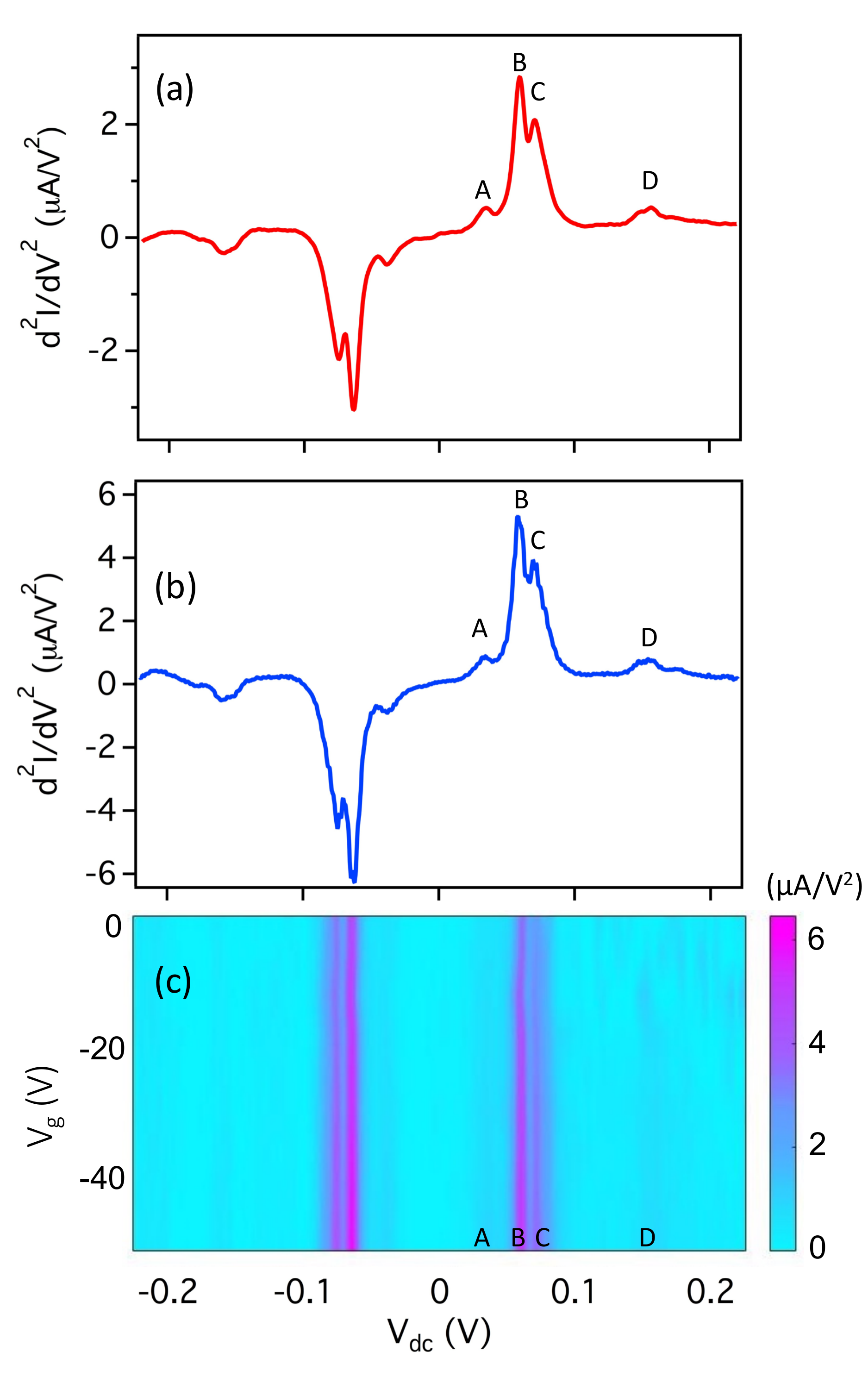}
\caption{(a) Derivative of conductance ($d^2I/dV^2$) as a function of source-drain bias ($V_{dc}$) for a graphite-hBN-Cr/Au tunnel junction. Four phonon peaks are marked as A-D. (b) $d^2I/dV^2$ vs. $V_{dc}$ for a MLG-hBN-Cr/Au tunnel junction, averaged over the gate voltage range 0 to -50 V. (c) The absolute value of the $d^2I/dV^2$ as a function of $V_g$ and $V_{dc}$ for the MLG device in (b). The phonon peaks A-D are found to be gate voltage independent.} 
\end{figure}

Weak tunneling features which we attribute to defects in the hBN tunel barrier are seen in some junctions. Figure 5(a) shows the $dI/dV$ characteristic for a BLG-hBN-Ag tunnel junction on Si/SiO$_2$ substrate, for various gate voltages. While the strong, finite voltage enhancement is a robust feature of the tunneling conductance, weak additional peaks, marked by arrows in the figure, are observed to shift around with gate voltage.  Figure 5(b) presents a color map of the conductance as a function of $V_g$ and $V_{dc}$ showing that these features evolve as sharp lines crossing in a roughly diamond pattern.  Similar signatures were also observed in a MLG-hBN-Ag tunnel junction on Si/SiO$_2$ substrate as seen in Fig.5(c).  The average tunneling conductance for this junction was much larger than for the device used for Fig. 5(b), most likely owing to a thinner hBN barrier, but the numerous weak features in the vicinity of $V_{dc} = 0$ again exhibit a clear diamond pattern. Interestingly, while the gate voltage dependent features are seen as peaks in $dI/dV$ in Fig.5(a) and (b), they appear as dips in Fig.5(c).

\begin{figure}
\centering
\includegraphics[scale=0.4]{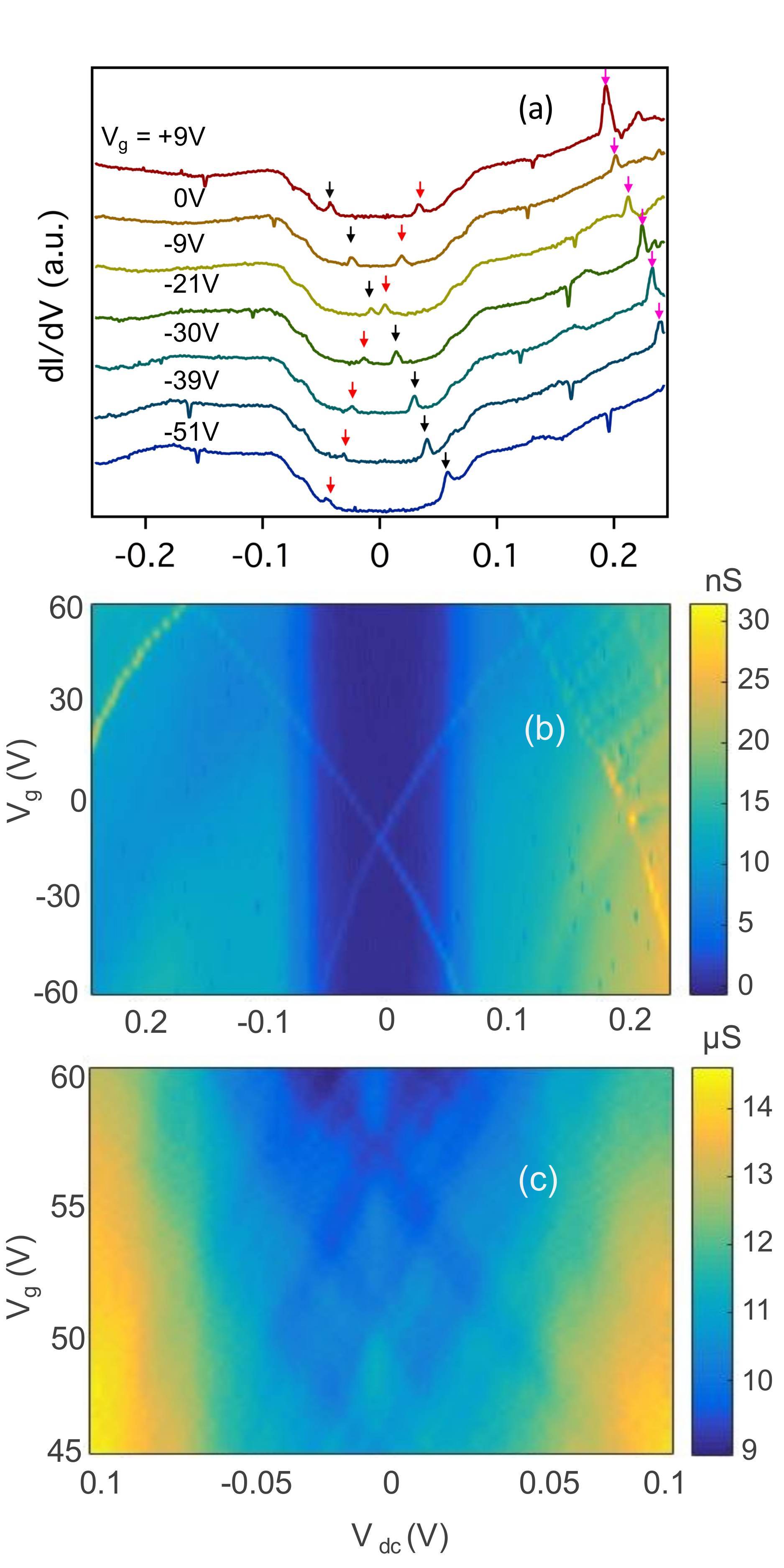}
\caption{(a) Conductance plots at various gate voltages for a BLG-hBN-Ag tunnel junction. The curves are offset for clarity. The arrows elucidate additional gate voltage-dependent features.  Color map of conductance plotted as a function of gate voltage and source-drain bias for (b) the BLG device in (a) and (c) a MLG-hBN-Ag tunnel junction.} 
\end{figure}

\section{Discussion}

We begin the discussion of these tunneling results with an idealized and oversimplified model.  If in-plane momentum is perfectly conserved, then the tunneling conductance at zero bias will vanish unless the Fermi surfaces (FS) of the two electrodes overlap in momentum space.  In graphene, the FS consists of tiny circles located at the $K$ and $K'$ points of the hexagonal Brillouin zone.  These points lie $1.70 \times 10^{10}$ m$^{-1}$ away from the zone center.  In contrast, for an idealized free electron metal the FS is a sphere at the zone center.   For silver, the radius of this sphere is $k_{F,Ag} \approx 1.2 \times 10^{10}$ m$^{-1}$.  Hence, in this approximation, there is no overlap between the graphene and silver Fermi surfaces and thus the zero bias tunneling conductance vanishes (at low temperature).

\subsection{Zero bias tunneling}
As Fig. 3 clearly shows, our tunnel junctions do in fact exhibit a non-zero, if small, tunneling conductance at zero bias.  The fact that in our MLG-hBN-Cr/Au junctions the average low bias conductance depends on the linear density of states of the graphene electrode demonstrates that this conductance is dominated by tunneling, not some parasitic conducting channel between the electrodes.  Although we do not know for certain from where this low bias tunneling conductance originates, there are a variety of possibilities.  For example, the Fermi surfaces of the metal and graphene/graphite electrodes may in fact overlap.  While this seems unlikely in the case where the metal electrode is Ag, the Cr/Au case is less clear.  For those junctions, the 5 nm Cr layer underlying the Au is most relevant.  The FS of Cr is highly complex and very far from a free electron sphere~\cite{Fawcett_RMP}.  Although the large electron pocket at the Cr zone center does not overlap with the $K$ and $K'$ points of graphene, the hole pockets around the vertices of the Cr Brillouin zone can.  The degree of this overlap is hard to assess, given the polycrystalline nature of the Cr layer (See supplementary materials for a discussion on the fermi surfaces of Cr, Ag and graphene).  Another speculative possibility is that intervalley scattering within the graphene may relax momentum conservation in the tunneling process sufficiently to overcome the lack of overlap between the graphene and metal Fermi surfaces.  Relatively rapid intervalley scattering rates have been recently reported in weak localization experiments on graphene~\cite{Chandni_PRB}.

\subsection{Phonon-assisted tunneling}
At finite bias, tunneling via various inelastic processes can occur~\cite{Zhang_nphys}.  In particular, phonons can provide the momentum needed to overcome the FS mismatch and allow tunneling to proceed.  The phonon spectra of graphene, graphite, and hBN are all quite similar due to their closely matched crystal structures~\cite{Yan_PRB,Serrano_PRL,Mohr_PRB}.  The energies of the various acoustic and optical phonons in these materials are all in the range below about 200 meV and are thus obvious candidates for explaining the various peaks in $d^2I/dV^2$ shown in Fig. 4.  Moreover, since the electronic momentum component perpendicular to the tunneling plane is not conserved, there is considerable phase space for in-plane phonons to link the three-dimensional FS of the metal with the two-dimensional FS of the graphene.  

A quantitative analysis of the kinematics of phonon-assisted tunneling requires a detailed knowledge of the electrode Fermi surfaces.  While the FS of graphene (and graphite) are well approximated as tiny circles at the corners of their Brillouin zones, the FS of the metal electrodes is not so simple.  This is particularly true for Cr whose Fermi surface is very complex~\cite{Fawcett_RMP}.  For this reason, we confine our analysis to a search for those energies at which the phonon density of states is particularly large. 

The phonon dispersion in hBN shows a flat band, and thus a high phonon density of states, along a line connecting the $M$ and $K$ points in the Brillouin zone at $\sim 40$ meV~\cite{Serrano_PRL}; this is close to peak A shown in the $d^2I/dV^2$ spectra shown in Fig. 4.  Similarly, multiple flat bands/crossings are present in the phonon dispersions of both hBN and graphene in the energy range $\sim 60-75$ meV~\cite{Yan_PRB,Serrano_PRL} where we observe the two strongest peaks, B and C, in $d^2I/dV^2$.  Around $150-170$ meV, where peak D resides, both materials again possess a large phonon density of states.  A more detailed
discussion of phonon-assisted tunneling in our devices can be found in the supplementary material. 

\subsection{Defect-assisted tunneling}
Although the hBN layers used in van der Waals heterostructures are often assumed to be pristine defect-free single crystals, this is not always the case.  Recent tunneling experiments on metal-hBN-metal (M-hBN-M) junctions revealed that is some cases the $IV$ curve exhibits a staircase-like pattern, very reminiscent of the Coulomb staircase observed in tunneling through a quantum dot~\cite{Chandni_NL}.  This analogy suggests that tunneling in M-hBN-M junctions can proceed via a two-step process involving tunneling from one metal electrode to a defect in the hBN and then from the defect to the other metal electrode.  From the voltage scale of the steps in the $IV$ curve the size of the defect was estimated to be in the few nanometer range, similar to that found in STM experiments which directly imaged defects in thin hBN layers~\cite{Wong_nnano}.  In such M-hBN-M devices, the metal electrodes very effectively screen out external gate electric fields. As a result, the classic diamond pattern was not observed.

In graphene-hBN-metal tunnel junctions the graphene lower electrode has such a low density of states that it cannot entirely screen an electric field produced by the underlying doped Si gate.  This imperfect screening, often referred to as a ``quantum capacitance'' effect, has been widely observed, both in graphene~\cite{Xia_nnano} and semiconductor heterostructures~\cite{Luryi_APL}.  In the present experiments, that portion of the gate electric field which penetrates through the graphene layer electrostatically shifts the energy of charged defects in the hBN layer and allows the diamond pattern to be observed. 

We speculate that the dips in Fig.5(c) might reflect Fano resonances~\cite{Gores_PRB}. Interference of a resonant state with a continuum (non-resonant background) leads to Fano line shapes in various nanostructures. When the coupling to the leads is strong and the nonresonant contribution to the conductance is of the same order as the resonant component, the line shapes of the conductance as a function of $V_g$ become unusual. When the nonresonant contribution dominates, symmetric dips are obtained in conductance vs $V_g$, in place of standard Coulomb oscillation peaks. In Fig.5(c), the background nonresonant component dominates as is evident from the large conductance ($\sim\!\mu$S). In Fig5.(a) and (b), the nonresonant component is much lower ($\sim$ nS) and we obtain Coulomb peaks and the usual diamond pattern.

\section{Conclusion}
In summary, tunneling spectroscopy measurements on metal-hBN-graphene (or graphite) junctions reveal clear signatures of both phonon-assisted and defect-assisted tunneling processes.   Phonon-assisted tunneling is particularly prominent in these devices owing to the small, and in some cases vanishing, overlap between the Fermi surfaces of the metal and the graphene (or graphite) electrodes. This lack of overlap sharply limits the phase space for direct, momentum conserving, electron tunneling.  Four prominent peaks in the second derivative of tunneling current, $d^2I/dV^2$, are associated with known phonons in the van der Waals heterostructure.  In addition, disorder, including point-like defects, in the hBN barrier layer enable tunneling events which do not conserve in-plane momentum.  In some devices, weak peaks in the tunneling conductance were found to present a diamond pattern when examined as a function of back gate voltage.  This confirms our prior conclusion that these peaks reflect single electron charging of nm-scale defects in the hBN layer.  

\begin{acknowledgements}
We thank S. Das Sarma, R. Sensarma and L. Zhao for useful discussions, and G. Rossman for the use of his Raman spectroscopy facility. Atomic force microscopy was done at the Molecular Materials Research Center of the Beckman Institute at the California Institute of Technology. This work was supported by the Institute for Quantum Information and Matter, an NSF Physics Frontiers Center with support of the Gordon and Betty Moore Foundation through Grant No. GBMF1250.
\end{acknowledgements}


\begin{references}
\bibitem{Smoliner} J. Smoliner, \textit{Tunnelling spectroscopy of low-dimensional states
}, Semicond. Sci. Technol. {\bf 11}, 1 (1996); E. L. Wolf, \textit{Principles of Electron Tunneling Spectroscopy}, (Oxford University Press, Oxford, 1985).
\bibitem{Reed} M. A. Reed et al., \textit{Observation of discrete electronic states in a zero-dimensional semiconductor nanostructure}, Phys. Rev. Lett. {\bf60}, 535 (1988).
\bibitem{JPE_PRB91} J.P. Eisenstein, T.J. Gramila, L.N. Pfeiffer, and K.W. West, \textit{Probing a two-dimensional Fermi surface by tunneling}, Phys. Rev. B. {\bf44}, 6511 (1991).
\bibitem{Gennser_PRL} U. Gennser et al., \textit{Probing band structure anisotropy in quantum wells via magnetotunneling}, Phys. Rev. Lett. {\bf67}, 3828 (1991).
\bibitem{JPE_PRL92} See, for example, J. P. Eisenstein, L. N. Pfeiffer, and K. W. West, \textit{Coulomb barrier to tunneling between parallel two-dimensional electron systems}, Phys. Rev. Lett. {\bf69}, 3804 (1992).
\bibitem{Chynoweth} A. G. Chynoweth, R. A. Logan, and D. E. Thomas, \textit{Phonon-Assisted Tunneling in Silicon and Germanium Esaki Junctions}, Phys. Rev. {\bf125}, 877 (1962).
\bibitem{Chandni_NL} U. Chandni, J. P. Eisenstein, K. Watanabe and T. Taniguchi, \textit{Evidence for Defect-Mediated Tunneling in Hexagonal Boron Nitride-Based Junctions}, Nano Lett. {\bf15}, 7329 (2015).
\bibitem{Chen_STM} C. J. Chen, \textit{Introduction to Scanning Tunneling Microscopy}, (Oxford University Press, Oxford, 2008).
\bibitem{Smoliner_PRL} J. Smoliner, W. Demmerle, G. Berthold, E. Gornik, and G. Weimann, \textit{Momentum conservation in tunneling processes between barrier-separated 2D-electron-gas systems}, Phys. Rev. Lett. {\bf63}, 2116 (1989).
\bibitem{Britnell_Science} L. Britnell et al., \textit{Field-effect tunneling transistor based on vertical graphene heterostructures}, Science {\bf335}, 947 (2012); F. Amet et al., \textit{
Tunneling spectroscopy of graphene-boron-nitride heterostructures}, Phys. Rev. B. {\bf85}, 073405 (2012).
\bibitem{Wang_Science} L. Wang et al., \textit{One-dimensional electrical contact to a two-dimensional material}, Science {\bf342}, 614 (2013).
\bibitem{Lee_APL} G-H. Lee et al., \textit{Electron tunneling through atomically flat and ultrathin hexagonal boron nitride}, Appl. Phys. Lett. {\bf99}, 243114 (2011).
\bibitem{Mishchenko_nnano} A. Mishchenko et al., \textit{Twist-controlled resonant tunnelling in graphene/boron nitride/graphene heterostructures}, Nature Nanotech. {\bf9}, 808 (2014); J. R. Wallbank et al., \textit{Tuning the valley and chiral quantum state of Dirac electrons in van der Waals heterostructures}, Science {\bf353}, 575 (2016); B. Fallahazad et al., \textit{Gate-Tunable Resonant Tunneling in Double Bilayer Graphene Heterostructures}, Nano Lett. {\bf15}, 428 (2014).
\bibitem{Fawcett_RMP} E. Fawcett, \textit{Spin-density-wave antiferromagnetism in chromium}, Rev. Mod. Phys. {\bf60}, 209 (1988).
\bibitem{Chandni_PRB} U. Chandni, E. A. Henriksen and J. P. Eisenstein, \textit{Transport in indium-decorated graphene}, Phys. Rev. B. {\bf91}, 245402 (2015).
\bibitem{Zhang_nphys} Y. Zhang et al.,\textit{Giant phonon-induced conductance in scanning tunnelling spectroscopy of gate-tunable graphene}, Nature Phys. {\bf4}, 627 (2008); F. D. Natterer et al., \textit{Strong Asymmetric Charge Carrier Dependence in Inelastic Electron Tunneling Spectroscopy of Graphene Phonons}, Phys. Rev. Lett. {\bf114}, 245502 (2015); S. Jung et al., \textit{Vibrational Properties of h-BN and h-BN-Graphene Heterostructures Probed by Inelastic Electron Tunneling Spectroscopy}, Scientific Rep. B. {\bf5}, 16642 (2015); E. E. Vdovin et al., \textit{Phonon-Assisted Resonant Tunneling of Electrons in Graphene–Boron Nitride Transistors}, Phys. Rev. Lett. {\bf116}, 186603 (2016).
\bibitem{Yan_PRB} J-A. Yan, W. Y. Ruan and M. Y. Chou, \textit{Phonon dispersions and vibrational properties of monolayer, bilayer, and trilayer graphene: Density-functional perturbation theory}, Phys. Rev. B. {\bf77}, 125401 (2008).
\bibitem{Serrano_PRL} J. Serrano et al., \textit{Vibrational Properties of Hexagonal Boron Nitride: Inelastic X-Ray Scattering and Ab Initio Calculations}, Phys. Rev. Lett. {\bf98}, 095503 (2007).
\bibitem{Mohr_PRB} M. Mohr et al., \textit{Phonon dispersion of graphite by inelastic x-ray scattering}, Phys. Rev. B. {\bf76}, 035439 (2007).
\bibitem{Wong_nnano} D. Wong et al., \textit{Characterization and manipulation of individual defects in insulating hexagonal boron nitride using scanning tunnelling microscopy}, Nature Nanotech. {\bf10}, 949 (2015).
\bibitem{Xia_nnano} J. Xia, F. Chen, J. Li and N. Tao, \textit{Measurement of the quantum capacitance of graphene}, Nature Nanotech. {\bf4}, 505 (2009).
\bibitem{Luryi_APL} S. Luryi, \textit{Quantum capacitance devices}, Appl. Phys. Lett. {\bf52}, 501 (1988).
\bibitem{Gores_PRB} J. Gores et al., \textit{Fano resonances in electronic transport through a single-electron transistor}, Phys. Rev. B. {\bf62}, 2188 (2000); W. Liang et al., \textit{Fabry - Perot interference in a nanotube electron waveguide}, Nature {\bf411}, 665 (2001).

\end{references}
\end{document}


\title{Signatures of phonon and defect-assisted tunneling in planar metal-hexagonal boron nitride-graphene junctions}
\centerline {\bf{\large SUPPLEMENTARY MATERIAL }}

\author{U. Chandni}
\email{chandniu@gmail.com}
\address{Institute for Quantum Information and Matter, Department of Physics, California Institute of Technology, 1200 E. California Blvd., Pasadena, California 91125, USA}
\author{K. Watanabe}
\address{National Institute for Materials Science, 1-1 Namiki Tsukuba, Ibaraki 305-0044, Japan}
 \author{T. Taniguchi}
\address{National Institute for Materials Science, 1-1 Namiki Tsukuba, Ibaraki 305-0044, Japan}
\author{J. P. Eisenstein}
\address{Institute for Quantum Information and Matter, Department of Physics, California Institute of Technology, 1200 E. California Blvd., Pasadena, California 91125, USA}

\maketitle

\subsection{\large S1. Role of metal Fermi surfaces in momentum conserved tunneling}
\begin{spacing}{1.4}

In-plane momentum is treated as a conserved quantity in the tunneling processes described in the main text. Here we discuss the silver and chromium Fermi surfaces (FS) and their contributions to the tunnel current. For simple metals, the FS can be treated as a sphere, which works reasonably well for the polycrystalline Ag tunnel electrodes. Fig.S1a shows a top view of the Brillouin zone (BZ) of graphene, where the tiny red circles at the corner $K$ and $K'$-points indicate the FS for a nominal density of  5 x 10$^{12}$cm$^{-2}$. The $\Gamma-K$ distance for graphene is $1.70 \times 10^{10}$ m$^{-1}$. The FS of Ag is shown as a sphere centered at the $\Gamma$-point ( $k_{F,Ag} \approx 1.2 \times 10^{10}$ m$^{-1}$) \cite{Ashcroft}. At low temperatures, the electrons that contribute to the tunnel current lie close to the Fermi surfaces of the metal and the graphene. Under normal circumstances, we observe that the Fermi surfaces have no overlap in momentum and hence elastic tunneling from the metal to the graphene layer is prohibited. Hence, only phonons or defect states in the heterostructure can contribute to inelastic tunneling processes, which are discussed in the following section.

\begin{figure*}
\centering
\includegraphics[scale=0.4]{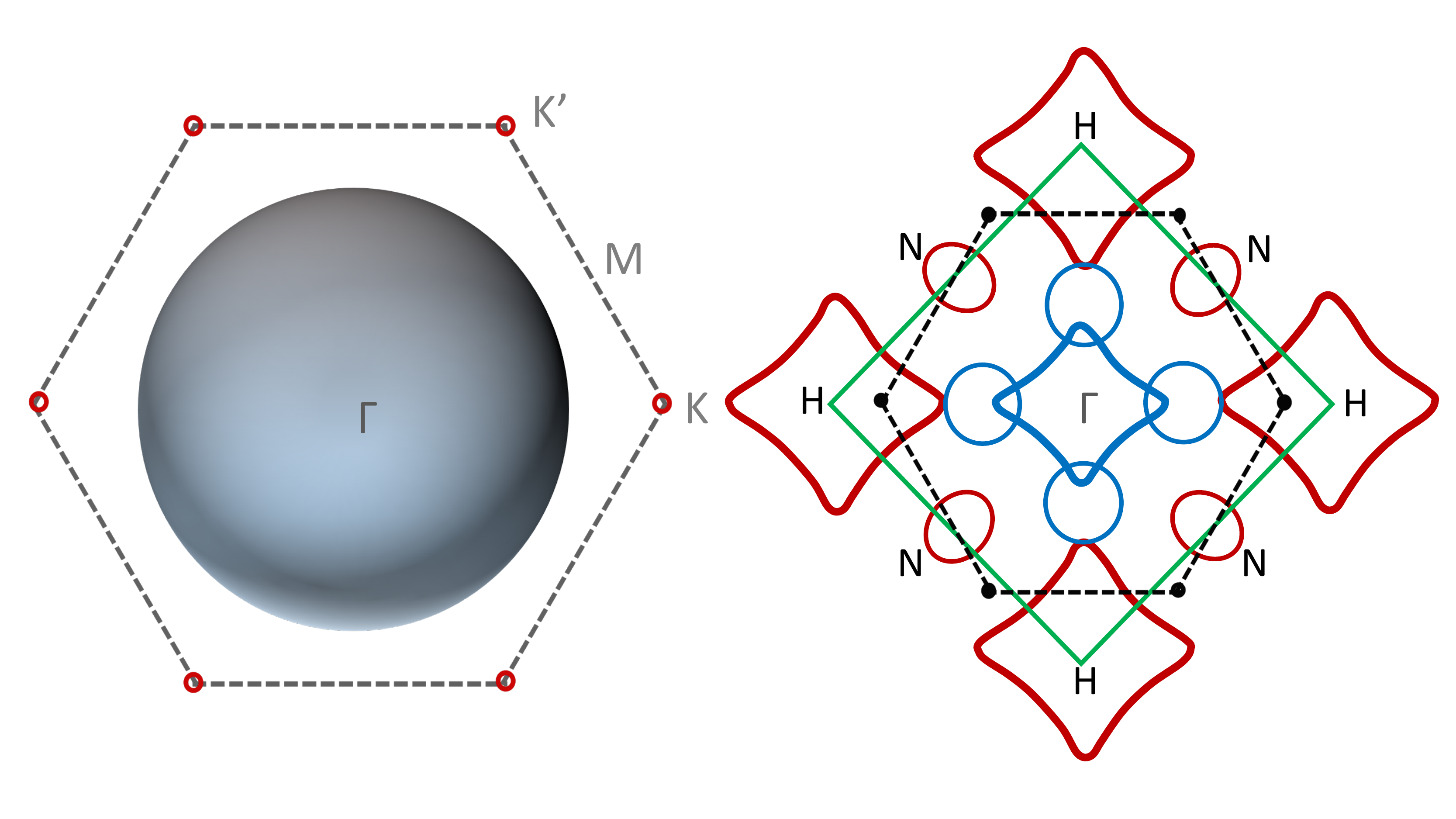}
\caption{(a) Top-view of the heterostructure elements as seen in the momentum space. The circles at the $K/K'$ points are the Fermi surfaces of graphene for a density of $n$ = 5 x $10^{12}$cm$^{-2}$. The nearly spherical metal Fermi surface is centered at the $\Gamma-$ point. (b) Chromium Fermi surface cross section in the (100) plane \cite{Fawcett_RMP}. The green square is the first BZ. The electron and hole octahedra are denoted by blue and red lines respectively. The black dashed line shows the graphene BZ and the tiny black dots indicate the graphene FS at the $K$ and $K'$-points. In this orientation a $K$ and a $K'$ point overlap the Cr FS.} 
\end{figure*}

In contrast, the FS of chromium is far from a simple sphere. The BZ of chromium is a dodecahedron and the FS has multiple electron and hole pockets within it \cite{Fawcett_RMP}. Fig. S1b shows a schematic of the FS of chromium \cite{Fawcett_RMP}.  At the zone center, there exists a non-spherical electron pocket shown in blue. Hole pockets (red lines) are seen at the face-centers ($N-$points) and corner points ($H-$points). The polycrystallinity of the top-Cr tunnel electrode results in an uncertainity in the orientation of this FS with respect to the graphene BZ (dotted black lines). Assuming for simplicity that the (100) surface is normal to the graphene BZ, we observe that the BZ of Cr is a square (green) superimposed on the hexagonal graphene BZ as seen in Fig.S1b. Clearly, portions of the Cr FS do fall outside the graphene BZ. The in-plane orientation of the two Fermi surfaces depends on the  orientations of the Cr crystallites. This presents a complicated phase space for momentum-conserved tunneling as the overlap of the two Fermi surfaces is governed by the relative orientations between the Cr crystallites and the graphene layer.

\subsection{\large S2. Phonon-assisted tunneling}
\linespread{1.6}
We now discuss the phonon-assisted tunneling phenomena assuming a spherical metal FS for simplicity. The electrons on the boundary of the spherical metal FS and the graphene FS contribute to tunneling. In Fig. S2, we schematically depict some of the possible phonon contributions to tunneling considering `in-plane' momentum conservation. An electron from the north or south pole of the metal Fermi surface (blue disks) can use its out-of-plane momentum (blue dotted line) which is not conserved and a $K$-point phonon mode (blue solid line) to tunnel on to the graphene FS. In addition an $M$-phonon mode (yellow solid line) may also contribute to the tunnel current since it can be shifted across the Brillouin zone to the K or $K'$-point (green line). If there exists an electron in the Fermi surface that has the right $z-$momentum (green dashed line), this transition can be aided. This suggests that several phonons can contribute to tunneling by supplying appropriate $z$-momenta to the elecrons on the surface of the metal FS. 

\begin{figure*}[h]
\centering
\includegraphics[scale=0.3]{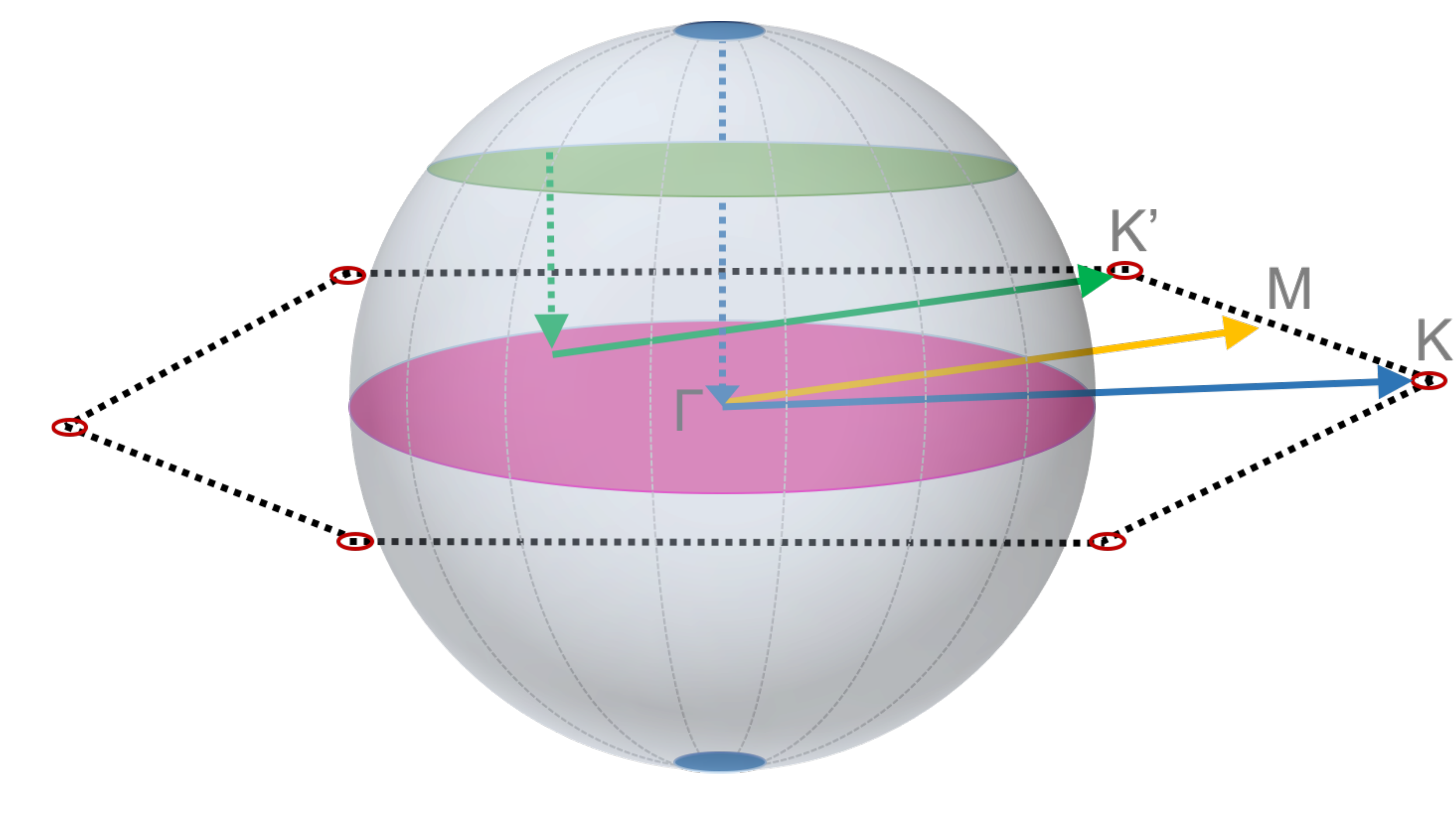}
\caption{Side-view of the Fermi surfaces of graphene and metal showing two possible phonon modes that contribute to tunneling.} 
\end{figure*}

Both graphene(/graphite) and h-BN have very similar phonon characteristics due to their closely matched crystal structure \cite{Yan_PRB,Serrano_PRL,Mohr_PRB}. Although the number of atoms in a unit cell is two for graphene and four for graphite (or hBN), many modes are doubly degenerate in graphite (or hBN) and hence the phonon dispersions are nearly identical with three optical (TO, LO or ZO) and three acoustic (TA, LA and ZA) branches. The phonons with the maximum density of states (with a flat band or a crossing in the phonon dispersion curve) will show the maximum contribution.  Interestingly, the above picture suggests that a relative orientation between the graphene and hBN layers will be inconsequential to the observed phonon modes as long as there are enough electrons in the Fermi surface that have corresponding $z$-momenta. The first peak in $d^2I/dV^2$ at $A = $36 meV can be attributed to the flat band in hBN in the $M-K$ direction arising from the ZA mode \cite{Serrano_PRL}. We observe that  the contributions to the two strongest peaks (at $B=61$ meV and $C=$ 74 meV respectively) come mainly from the ZA and ZO modes of the graphene and hBN \cite{Yan_PRB,Serrano_PRL}. This is in agreement with previous observations that the coupling of the out of plane modes to the graphene wavefunctions aid tunneling significantly \cite{Zhang_nphys}. More specifically, hBN has a flat band around $\sim 75$ meV in the $M-K$ direction arising from its ZO mode \cite{Serrano_PRL}. Graphene has a flat band at the $M-$point arising from the ZA mode, a crossing at the $K-$ point arising from its ZA/ZO modes at $\sim$ 65 meV and a ZA/ZO/TA band crossing at the $M-$point at $\sim $ 80 meV \cite{Yan_PRB}. Around the range $\sim 150-170$ meV, both hBN and graphene have many flat bands and crossings from LO and TO modes, which possibly lead to the broad peak that we observe at $D= $166 meV.


\end{spacing}